# From Single Spins to Ensembles: A Calibrated Symbolic Regression Pipeline for NV Spectroscopy

Evan Zhang[1,2], Zhenghang Zhi[3], Zewei Wang[3], and Zeheng Wang[4,†]
[1] School of Physics, UNSW, Sydney, Australia; [2] School of Electrical Engineering and Telecommunications, UNSW; [3] Cryo-X Tech Co., Shanghai, China; [4] Hangzhou Institute of Technology, Xidian University, Hangzhou, China
[†] Email: zenwang@outlook.com

Characterizing nitrogen-vacancy (NV) ensembles requires connecting their collective spectra to individual centers that experience different local environments. We develop a symbolic-regression pipeline that carries single-center spin structure into a calibrated ensemble model by generating individual spectral responses and combining them with explicit weights. A constrained search selects Hermitian operator terms from single-center optically detected magnetic resonance spectra, fits their coefficients, and retains the selected expressions as source models for the ensemble. Ensemble calibration then accounts for resonance shifts, local disorder, orientation, and optical response while keeping the individual spectral contributions accessible. On public single-center and ensemble data, the zero-field ensemble held-frequency root-mean-square error decreases from 0.008835 to 0.003872 in normalized photoluminescence, a 56.2% improvement over matched uncorrected composition. In controlled QuTiP ensembles of 128 sites spanning four NV orientations, models calibrated at 0 and 3 mT predict an independently held-out 5.17 mT spectrum with errors of 0.000207–0.000228 across three random seeds, comparable to the 0.000200 noise standard deviation. Resolving the simulated response into its individual-center contributions shows how local spectral diversity combines into the ensemble line shape. By retaining the source Hamiltonians while calibrating sample-dependent spectral differences, the pipeline connects local spin physics to collective optical readout and makes the contributions to an ensemble line shape available for physical interpretation.

## I. INTRODUCTION

Nitrogen-vacancy (NV) centers in diamond offer optically accessible spins for magnetic sensing and imaging [1–3]. High-resolution single-center spectroscopy resolves the local spin structure underlying this response, whereas ensemble spectroscopy combines centers with different orientations and environments [4–7]. Their overlapping resonances make the relation between local spin structure and the measured ensemble line shape central to quantitative interpretation. Quantum diamond microscopy and vector magnetometry therefore benefit from spectral models that retain this relation while accounting for ensemble broadening and optical response [8,9]. The information available at the local level is illustrated by single-NV nuclear magnetic resonance with chemical resolution, demonstrated by Aslam et al. [10]. The question addressed here is how to reuse the spin structure identified in single-center spectra while accounting for the local environments and optical response of a target ensemble.

The connection between an individual center and an ensemble begins with the composition of their measured responses. For independent centers with additive optical readout, the ensemble spectrum is a weighted sum of local spectra. Each local spectrum can depend nonlinearly on Hamiltonian parameters, even though the optical signals add linearly. The environmental sensitivity underlying this local description also appears in single-NV relaxometry, where changes in the surrounding molecules alter the measured spin response [11]. In the static ODMR setting considered here, a distribution of local parameters represents spectral heterogeneity, and orientation-dependent projections connect each center to the applied field. Retaining these contributions separately makes it possible to examine how a change in a local resonance, a population weight, or a disorder distribution propagates into the collective spectrum.

Established NV theory supplies a physically meaningful operator vocabulary for this construction. Ground-state fine structure, nitrogen hyperfine coupling, nearby nuclear spins, and magnetic-field terms can be represented using a small set of spin operators [12–14]. High-resolution zero-field spectroscopy reveals additional local spectral structure [15], while transverse splitting, temperature-sensitive resonance positions, acquisition-dependent linewidths, and orientation-dependent polarization motivate explicit calibration and readout choices [16–20]. Ensemble dephasing studies further establish the importance of site-to-site environments [21,22]. These physical ingredients connect a compact local Hamiltonian to the spectral differences that an ensemble model must represent. We use centers A and B as two measured local prototypes, then calibrate their composition to a target ensemble acquired from a different specimen. This separation lets the source Hamiltonians serve as reusable local models, while target-specific calibration accounts for how their spectral contributions appear in the measured ensemble.

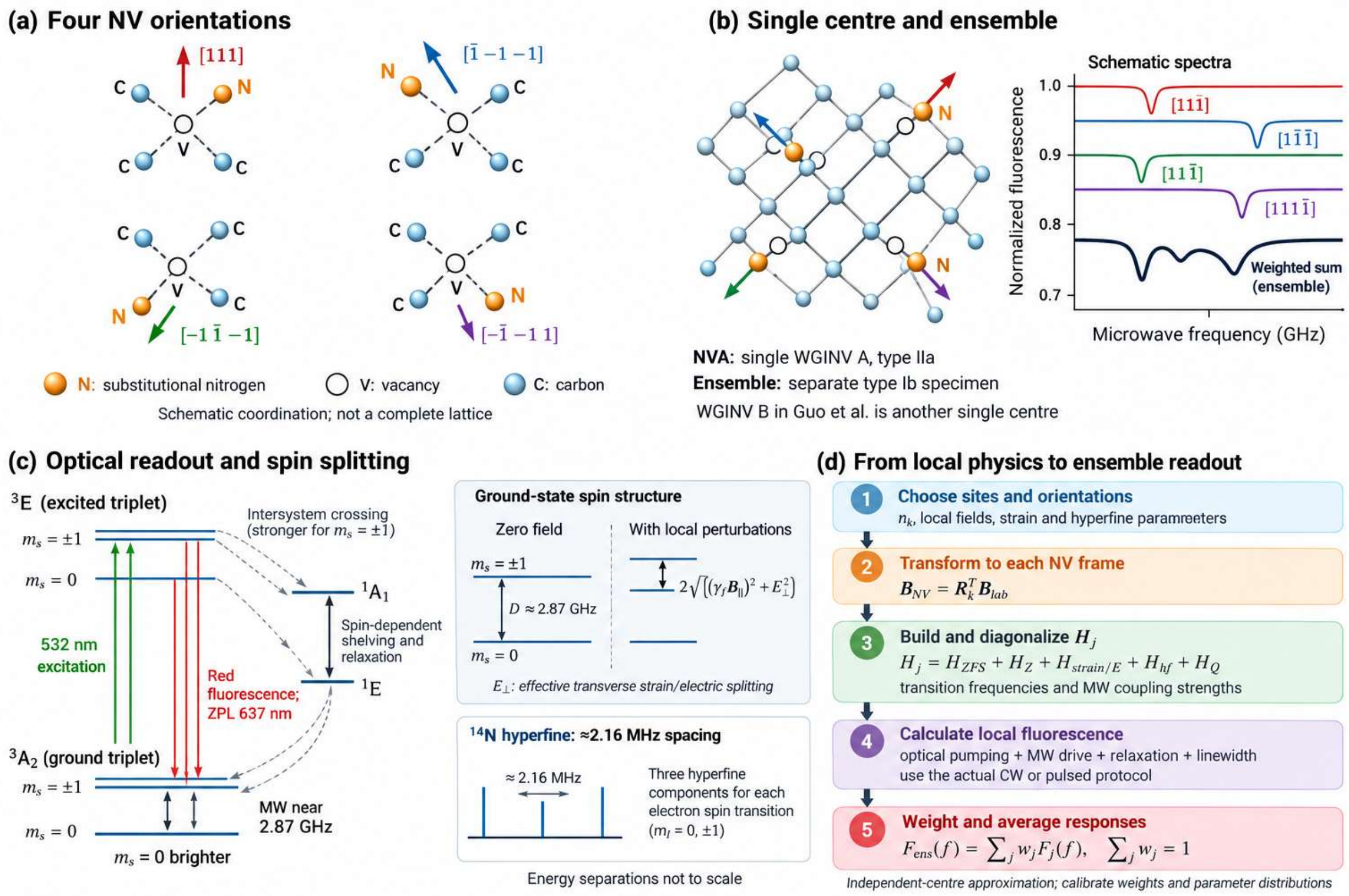

**FIG. 1.** Single-center structure and ensemble spectral composition. (a) Scripted rendering of an ideal diamond lattice with substitutional nitrogen (N), its nearest-neighbor vacancy (V), carbon atoms (C), and the NV symmetry axis along a ⟨111⟩ direction. The open cage denotes the vacant lattice site. (b) A schematic diamond fragment containing NV centers along the four tetrahedral orientations illustrates how local environments and orientation enter the ensemble description. (c) Ground-state spin-level diagram showing the zero-field splitting D and the local Zeeman, transverse-splitting, and hyperfine contributions. (d) Selected local operators, site parameters, and orientation transforms determine individual spectral responses, whose weighted sum gives the ensemble readout. Lattice fragments and level separations are schematic. The geometry was constructed and rendered in Blender with an AI-assisted script.

Hamiltonian learning provides several routes to extracting physical models from observations, including Bayesian parameter estimation, quantum-assisted inference, local-measurement reconstruction, and learned descriptions tested on NV platforms [23–27]. These approaches connect measured responses to model parameters or structures through specified inference procedures. Machine-learning models trained on continuous quantum measurements provide a further applied route [28]. Symbolic regression and sparse identification emphasize explicit mathematical expressions and controlled model complexity [29,30]. We apply that emphasis to the composition problem: select and fit a local Hamiltonian within a defined Hermitian operator library, retain the selected expression, and use it to generate the individual responses that build an ensemble spectrum. This formulation gives operator selection, parameter calibration, and spectral composition distinct roles within one characterization pipeline.

We implement this pipeline using public raw data for laser-written waveguide-integrated NV spins [31,32] and a separately implemented QuTiP simulation [33,34]. The workflow links operator selection, coefficient fitting, source-model freezing, ensemble calibration, and weighted spectral composition. Experimental evaluation withholds frequencies within the measured scans to test the calibrated line shapes. The numerical evaluation then withholds an entire magnetic-field condition and resolves the simulated ensemble into its individual-center spectra. Together, these tests connect the accuracy of the collective response to the explicit local contributions from which it is constructed.

## A. NV centers and optical spin readout

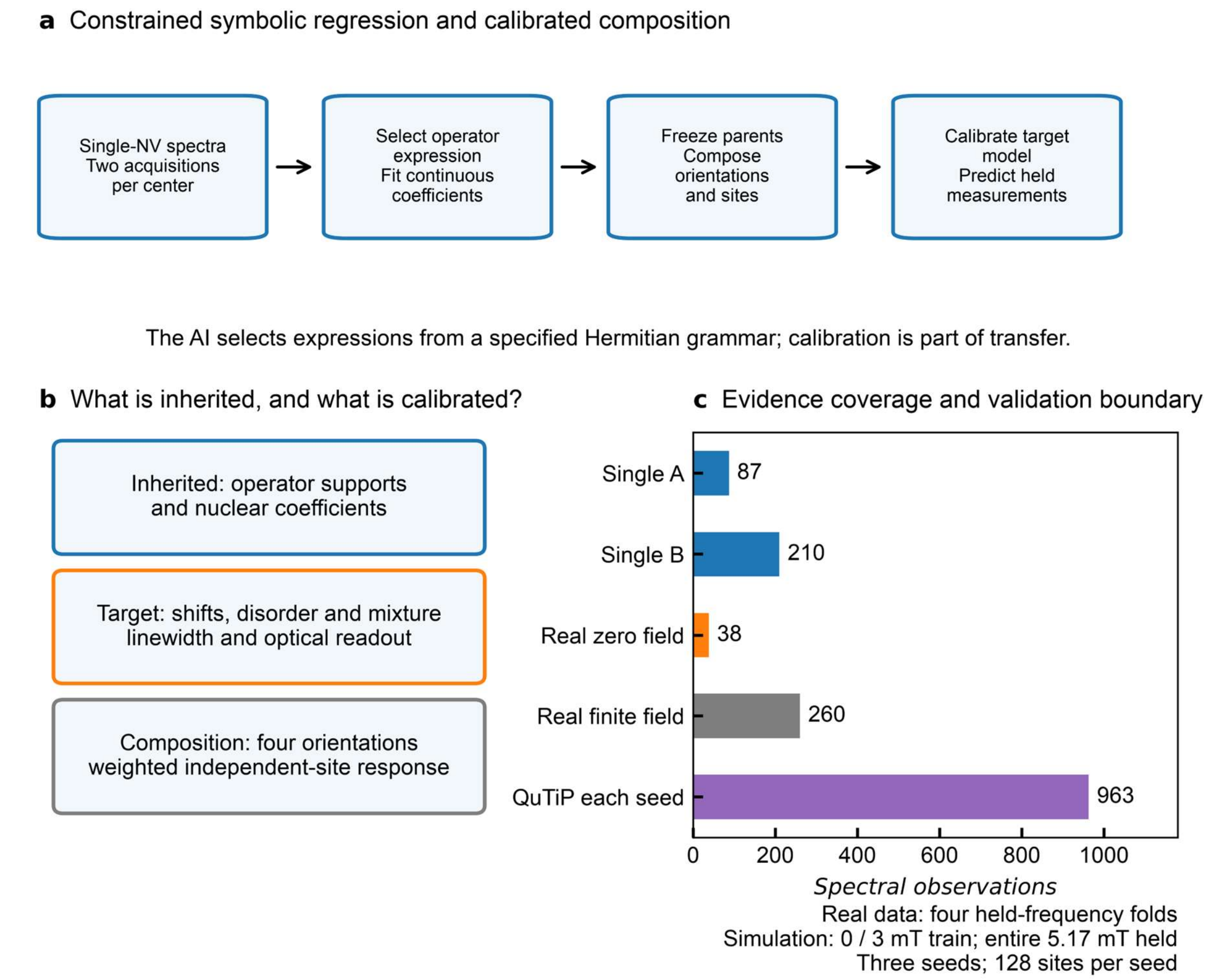

**FIG. 2.** Calibrated pathway from single NV centers to ensemble spectra. (a) Constrained symbolic regression selects operator expressions and fits continuous coefficients; the resulting local expressions are frozen before calibrated composition. (b) Source-specific supports and nuclear coefficients are inherited, while target shifts, disorder, mixture, and readout are calibrated. This separation retains the single-center model structure across different specimens. (c) Spectral observations in each experimental task and all three simulated conditions per seed. Center A contains 27 CW and 60 pulsed points; center B contains 31 CW and 179 pulsed points. Experimental prediction uses four cyclic held-frequency folds within the scans. Simulation training uses 0 and 3 mT, with the entire 5.17 mT condition withheld for each of three independently seeded 128-site populations. The source centers and experimental ensemble come from separate specimens; counts denote spectral observations.

A negatively charged nitrogen-vacancy center ($NV^{-}$) consists of a substitutional nitrogen atom adjacent to a carbon vacancy [Fig. 1(a)]. The electronic ground state is a spin triplet ($S = 1$) with projections $m_s = 0, \pm 1$. Its zero-field splitting is approximately 2.87 GHz, with additional structure produced by magnetic fields, transverse electric or strain effects, and hyperfine interactions. These contributions define the effective ground-state spin Hamiltonian used to calculate each local spectrum [Fig. 1(c)].

## B. Experimental data and sample distinction

We analyze spectra reported by Guo et al. for laser-written waveguide-integrated NV centers [31,32]. Isolated centers were characterized in low-nitrogen type-IIa diamond (sample NVA), while ensemble spectra were acquired in nitrogen-rich type-Ib diamond (sample NVB). The distinct specimens provide the source and target of the transfer: the single-center data determine local operator representations, and the ensemble data determine the corrections needed for their collective response
Each sample is distinguished by its confocal photoluminescence (PL) maps in both the $x$-$y$ and $x$-$z$ planes (Fig. 4). The waveguide-integrated nitrogen-vacancy centers (WGINVs) were fabricated by femtosecond laser writing followed by thermal annealing. Parallel modification lines separated by 13 μm define the optical waveguides.

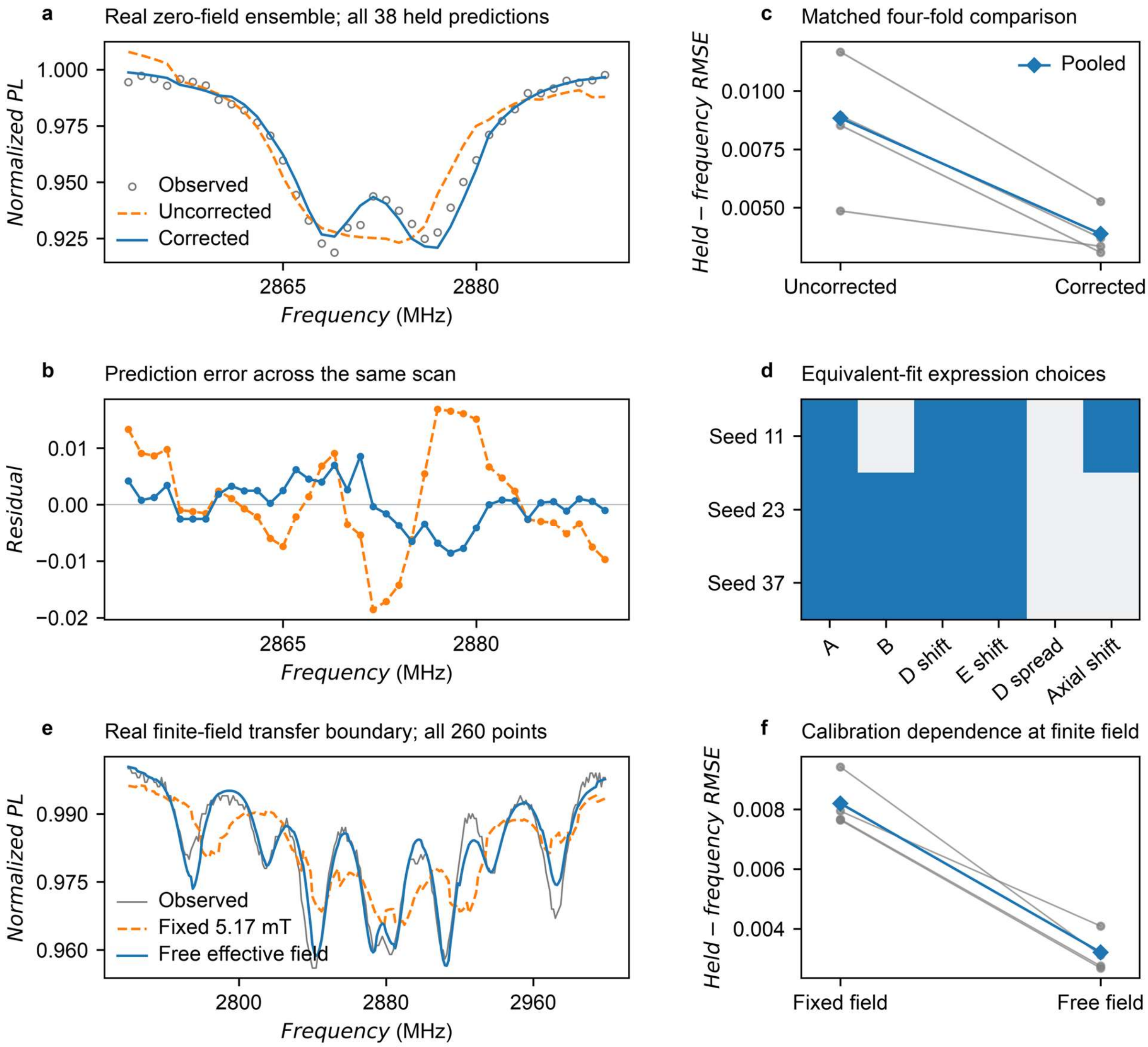

**FIG. 3.** Experimental ensemble calibration at zero and finite field. (a,b) All 38 zero-field observations, four-fold held predictions from corrected and uncorrected composition, and their residuals. Both procedures refit linewidth and identical optical readout terms on the target training points. (c) Matched fold and pooled RMSEs; correction reduces the pooled error from 0.008835 to 0.003872, or 56.2%. (d) Full-data target supports for three search seeds, showing the selected parents and four available corrections. (e,f) All 260 finite-field observations, held predictions, and matched fold errors with the field magnitude fixed at the reported 5.17 mT or fitted as an effective parameter. Pooled errors are 0.008199 and 0.003209, respectively; the full-data effective field is 3.76 mT. Experimental folds partition frequencies within each scan.

In low-nitrogen type-IIa diamond, additional low-energy static exposures create vacancies near the waveguide center, and annealing yields isolated single NVs WGINV A and WGINV B at different longitudinal positions along the waveguide. Conversely, in nitrogen-rich type-Ib diamond, vacancies generated within the laser-modified sidewalls diffuse during annealing and are captured by substitutional nitrogen, producing ensemble WGINVs.

Optically detected magnetic resonance (ODMR) connects spin-dependent fluorescence with transitions in the electronic ground state. Green optical excitation preferentially polarizes the $NV^-$ spin toward $m_s = 0$; spin-dependent intersystem crossing produces lower fluorescence for populations in the $m_s = \pm 1$ manifolds. A microwave field near a transition frequency changes those populations and yields an ODMR feature.

Continuous-wave (CW) ODMR applies optical and microwave excitation concurrently, whereas pulsed ODMR separates preparation, manipulation, and readout. The two protocols probe the same local ground-state structure under different drive and readout conditions. These conditions can change the linewidth, contrast, and background of the observed spectrum. Accordingly, CW and pulsed spectra of a given center share a fitted spin Hamiltonian, while their phenomenological linewidths and optical readout coefficients are fitted separately.

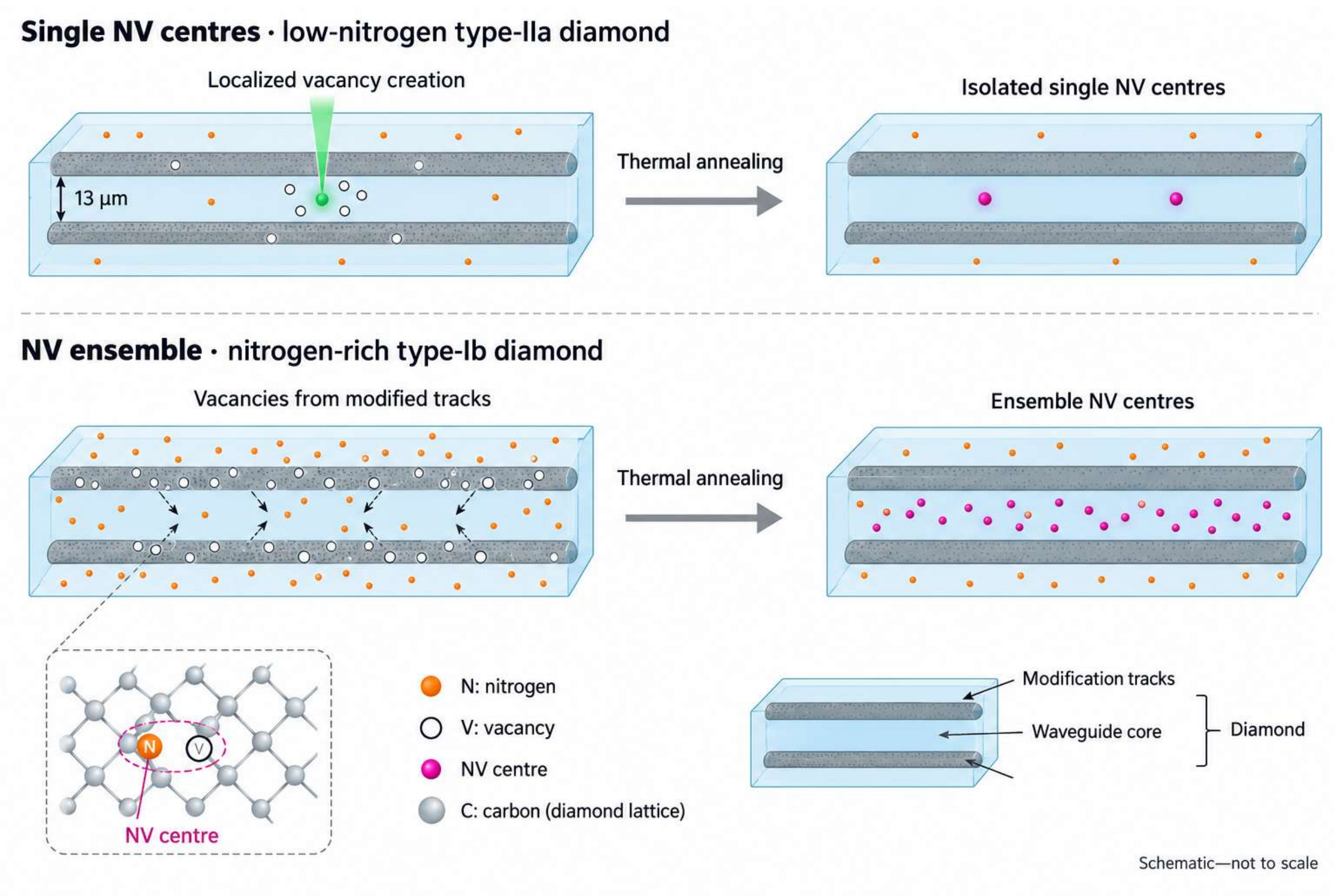

**FIG. 4. Fabrication of single and ensemble waveguide-integrated nitrogen-vacancy centres in diamond.** Two parallel laser-written modification tracks separated by 13 μm define the waveguide region. **Top row:** In low-nitrogen type-IIa diamond, additional localized laser exposure creates vacancies near the waveguide centre; Different thermal annealing exposure varied the quantity of isolated NV centres formed. **Bottom row:** In high-nitrogen type-Ib diamond, vacancies generated by laser exposure diffuse during annealing and are captured by substitutional nitrogen, producing an NV ensemble within the waveguide core. Orange markers represent nitrogen, open circles denote vacancies, and magenta markers indicate formed NV centres. The inset illustrates a substitutional nitrogen atom adjacent to a vacant lattice site. AI-assisted cartoon summarizes the fabrication description of Y. Guo *et al.*, *APL Photonics* **9**, 076103 (2024). Geometry and marker densities are not to scale.

At approximately zero applied field, center A displayed a CW splitting of 2.26 MHz and center B a splitting of 3.87 MHz. Pulsed measurements resolved four prominent features for A and eight for B; measurements near 4.6 mT resolved additional hyperfine structure. The two centers therefore provide distinct local spectra with which to test the constrained operator library.

We fit separate effective Hamiltonians to centers A and B. Within each center, the spin coefficients are shared across the CW and pulsed acquisitions, with the local environment treated as unchanged between measurements. Scan-dependent linewidth, contrast, offset, and background slope describe the corresponding measurement responses. Joint fitting therefore connects the resolved features in both acquisitions to one local operator expression.

The source fits supply two compact, inspectable spin representations. Their selected operator supports and nuclear-spin coefficients remain fixed during ensemble composition. Target shifts, splitting disorder, orientation contributions, and optical response then determine how these local prototypes contribute to the measured ensemble spectrum. Figure 1(b) illustrates the relation between local environments, individual responses, and collective readout.

## II. MODEL SELECTION AND ENSEMBLE COMPOSITION

Figure 2 summarizes the calibrated transfer workflow. Single-center CW and pulsed ODMR spectra constrain a common Hamiltonian for each defect within a six-term Hermitian operator library. Separate scan-dependent readout functions describe optical contrast and linewidth. A discrete search selects the retained terms, and bounded nonlinear least squares fits their coefficients. The resulting local expression supplies the eigenstates and transition frequencies used to calculate an individual spectrum. Appendixes A and B specify the basis, line-shape model, selection score, and search budget [35].

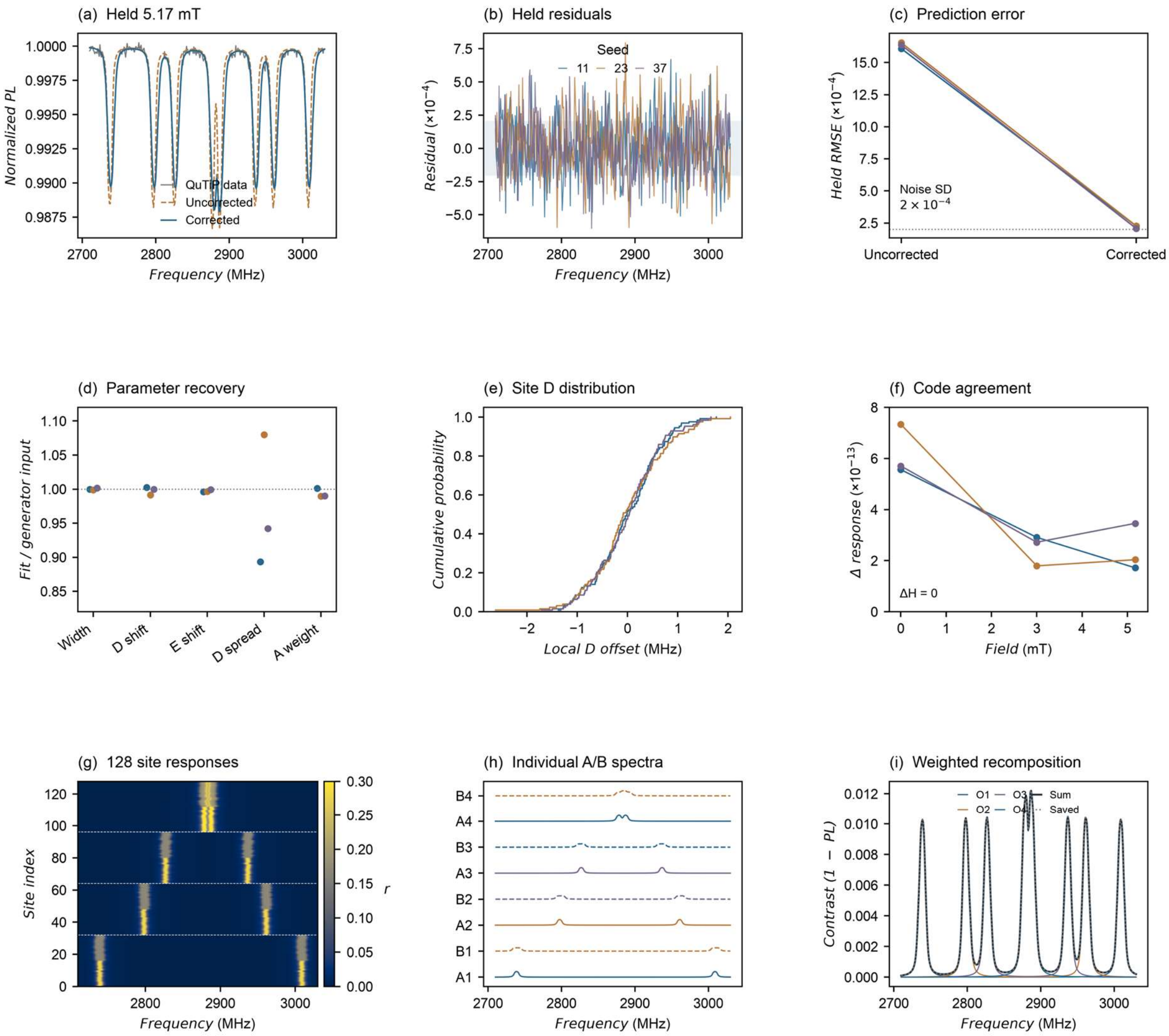

**FIG. 5.** Whole-field prediction and individual spectral composition in the QuTiP ensemble. (a) All 321 observations at the withheld 5.17 mT field for seed 11, with corrected and uncorrected predictions. (b) Held-field residuals for seeds 11, 23, and 37; shading denotes ±1 noise standard deviation (0.000200 normalized PL). (c) Paired held-field RMSEs and the noise scale. (d) Fitted parameters divided by generator inputs. (e) D-offset distributions for all 128 sites per seed. (f) Maximum QuTiP–NumPy response differences; retained local and pair Hamiltonian differences are zero. (g) Calculated local responses r for all 128 saved seed-11 sites, grouped by orientation in blocks of 32. (h) The first parent-A and parent-B site per orientation, with vertical offsets of 0.52; solid and dashed curves identify A and B. (i) Each orientation contributes 0.18 times the sum of its 32 local responses divided by 128. Their sum reproduces the saved noise-free ensemble contrast, $1 - \mathrm{PL}$, with maximum absolute difference below $10^{-12}$. Panels (g,h) show unweighted responses; (i) applies uniform site weights and optical contrast. Components use saved parameters of the independent-site simulation at 5.17 mT. Training uses 0 and 3 mT, with the entire 5.17 mT condition held out.

The full-data source models are frozen before ensemble fitting. Their selected operator supports and nuclear coefficients are inherited; target calibration can choose either parent or a mixture and fit mean resonance shifts, a distribution of zero-field splittings, and an effective axial term. Target linewidth and optical readout parameters are fitted independently. Linear optical coefficients are eliminated from the nonlinear optimization using variable projection [36,37]. The ensemble spectrum is then assembled from the calibrated local responses, preserving the relation between each retained Hamiltonian, its spectral contribution, and the collective line shape.

Model selection and continuous parameter fitting use only the training portion of each evaluation split. Four cyclic frequency-block folds are used for the experimental spectra, with identical partitions and optical-readout flexibility for corrected and uncorrected composition. The comparator also permits parent selection, mixture weights, and linewidth fitting. In the simulation, spectra at two magnetic fields train the model and a third field condition is reserved in full. These complementary splits test interpolation within measured scans and prediction under a changed field, respectively. Appendix C defines the experimental partitions.

## III. SINGLE-CENTER EXPRESSION FITTING

The constrained search fits the four measured single-center spectra and predicts frequencies withheld from each scan (Fig. 3). Center A contributes 87 observations (27 CW and 60 pulsed), and center B contributes 210 (31 CW and 179 pulsed). Their pooled held-frequency RMSEs are 0.01094 and 0.00932 in normalized photoluminescence, respectively. The corresponding all-six-term fits yield 0.01337 and 0.01362 using the same finite number of coefficient initializations. The spectra and residuals show how each selected expression reproduces its center's resolved features across both acquisition protocols. Fold-level supports are reported in Appendix C.

The full-data source expressions retain six terms for center A and five for center B (Fig. 3). Their different coefficients reproduce distinct local spectral patterns, providing two experimentally informed prototypes for ensemble composition. Selected supports vary among held-frequency folds under the finite expression-search and coefficient-optimization budgets. Appendix A gives the fitted coefficients, bounds, and operator conventions of the frozen source parents.

For ensemble composition, we freeze the full-data source expressions from the prespecified search seed 11. Both parent models are learned exclusively from single-center spectra, so their operator supports and nuclear couplings are determined before the target ensemble is fitted. Ensemble calibration then adjusts their collective response to the distinct target specimen. This sequence preserves a traceable connection from the individually measured spectra to the composed ensemble model.

## IV. EXPERIMENTAL ENSEMBLE CALIBRATION

The ensemble-specific corrections improve prediction of the zero-field target spectrum (Fig. 5). Calibration reduces the pooled held-frequency RMSE of the 38-point scan from 0.008835 for uncorrected composition to 0.003872 for corrected composition, a 56.2% reduction. The corrected model improves all four matched folds. Both procedures optimize identical linewidth and optical-readout parameters, so the comparison isolates the contribution of the added Hamiltonian corrections within the inherited source-model family.

The primary full-data fit selects parent A with corrections to the mean splitting, transverse coupling, and effective axial term. Additional search seeds select mixtures of the two parents with comparable fitting accuracy (Fig. 5 and Appendix D). These alternative effective parameterizations retain different local compositions while describing a similar aggregate line shape. Keeping the selected supports and mixture weights explicit makes that model variation directly inspectable.

The 260-point finite-field ensemble scan examines how field calibration affects the composed response. With the reported 5.17 mT field fixed, the inherited model yields a held-frequency RMSE of 0.008199. Fitting an effective field magnitude reduces the RMSE to 0.003209 and gives a full-data effective value of 3.76 mT (Fig. 5). The comparison shows that the field parameter materially changes the local transition positions and their aggregate spectrum. It consequently places field and sample characterization alongside local-model selection in the calibration procedure [38].

## V. CONTROLLED WHOLE-CONDITION TEST WITH QUTIP

To test prediction under a changed field, we generated three independent 128-site ensembles with QuTiP (Fig. 5). Each comprises 18-level local spin systems distributed equally among four tetrahedral NV orientations and two source parents. Site-to-site splitting offsets introduce local spectral variation. QuTiP supplies the local eigensystems for the simulated spectra, whereas the fitted ensemble model uses independently implemented NumPy operators and five-node Gaussian quadrature for the splitting distribution. The resolved site responses in Figs. 5(g)–6(i) show how orientation, parent identity, and local offsets generate distinct spectral contributions whose weighted sum reproduces the stored clean ensemble response. Appendix E gives the composition rule and population parameters.

The model is calibrated on spectra at 0 and 3 mT and predicts the complete 5.17 mT condition, containing 321 frequencies for each seed. Held-condition RMSEs of 0.000207–0.000228 approach the imposed noise standard deviation of 0.000200, compared with 0.00161–0.00165 for uncorrected composition.

Agreement at the withheld field tests the propagation of the calibrated local structure into a changed ensemble spectrum. QuTiP and NumPy implementations also give identical local and explicit two-site Hamiltonian checks, with line-response differences below $7.4 \times 10^{-13}$.

The full four-correction expression is included in the initialized candidate set and selected at the second expression evaluation for every seed. Its continuous coefficients are optimized using the two training conditions. The resulting prediction combines those calibrated coefficients with the local orientation transforms and spectral-composition rule. Resolving the simulated response by center provides a direct view of this construction, while the independent implementation checks verify its numerical consistency. Search histories, residuals, and reproduction checks are collected in Appendix F.

## VI. DISCUSSION

The central result is a composition procedure that carries single-center spin structure into a calibrated ensemble response while retaining the contributions from heterogeneous local environments. The learned object is an effective Hamiltonian: its selected spin operators, fitted coefficients, eigenstates, and transition frequencies remain available for inspection. We construct this object by selecting among six optional Hermitian terms, fitting the coefficients of each candidate, and ranking the resulting expressions with a complexity-penalized score. The discrete representation is a six-bit support mask, which gives every retained term a direct physical interpretation. This representation also determines how the object is reused. Once a source expression is frozen, its local spectrum can be evaluated under a specified field and parameter offset, then combined with other centers through the readout model. The spectral representation is therefore explicit at both levels: the local Hamiltonian determines individual transition structure, and the composition rule determines how that structure appears in the ensemble line shape. Figures 3 and 5 connect these two levels through measured source spectra and resolved simulated site responses.

The distinction between inherited local structure and target calibration explains the experimental improvement. The source models specify which spin terms generate the local transitions, while ensemble-specific shifts, disorder, and readout parameters determine where and with what weight those transitions appear in the target spectrum. Composition preserves these roles separately. A change in mean splitting shifts the contributing resonances; a distribution of splitting offsets spreads their positions; and a change in parent or orientation weight redistributes their contributions. These operations connect the microscopic form of the local model to the observed collective line shape. In the controlled ensemble, the individual-center spectra display this connection directly: centers with different orientations and offsets contribute different resonance patterns, and their weighted sum returns the ensemble response. An ensemble can thus be characterized through both its aggregate prediction and the modeled distribution of local contributions. This inspectable description is useful for organizing calibration of NV magnetometry and waveguide-integrated sensing, where field geometry and sample-dependent spectral structure enter the same measured response.

The pipeline brings Hamiltonian learning and symbolic model selection into a common composition procedure [25,27,29,30]. Its defining feature is the retention of an explicit physical object throughout the calculation. The selected Hamiltonian generates spectra at specified fields, the readout model converts those spectra into optical signals, and the ensemble rule combines the signals with stated weights. Each stage can therefore be examined or modified through a corresponding physical quantity. Bayesian estimation, neural inverse maps, and physics-informed optimization can supply different parameter-inference strategies around such a forward model; the present implementation uses bounded least squares and discrete support selection. The local operator expression also makes a concrete distinction between changing a coefficient and changing the allowed interaction structure. Coefficient changes remain within the selected model, whereas a revised operator library defines a new search space that can be tested with the same prediction procedure. In the controlled evaluation, QuTiP supplies finite-site spectra and the NumPy learner supplies a quadrature approximation to their distribution. Their separately constructed matrices and the withheld-field test examine whether these two implementations carry the same local physics consistently into an ensemble prediction.

The experimental spectra show why composition and calibration must be considered together. The two source centers were measured in type-IIa diamond, whereas the target ensemble was measured in type-Ib diamond. Transferring their operator representations therefore requires fitting the target's effective spectral shifts and optical response. On the zero-field scan, those corrections reduce the pooled held-frequency RMSE from 0.008835 to 0.003872 and improve all four matched folds. Because both procedures fit the same linewidth and optical coefficients, the improvement tracks the added Hamiltonian corrections within the inherited model family. The primary full-data solution uses parent A, while other search seeds retain mixtures with similar aggregate accuracy. The recorded alternatives show which aspects of the composition vary while the predicted ensemble response remains similar. At finite field, fixing the magnitude at 5.17 mT gives an RMSE of 0.008199, whereas fitting an effective magnitude gives 0.003209 and a full-data value of 3.76 mT. This sensitivity links the quality of the collective spectrum to the field projections entering the local Hamiltonians. Material and field characterization are consequently part of the calibration problem [38]. Recording both field treatments and the alternative compositions gives subsequent measurements concrete quantities to constrain: local resonance offsets, orientation weights, parent fractions, and the effective field configuration.

The controlled ensemble connects that calibration problem to prediction under a changed measurement condition. Across three independently seeded 128-site populations, fitting at 0 and 3 mT yields complete 5.17 mT spectra with RMSEs of 0.000207–0.000228, close to the specified noise scale and below the 0.00161–0.00165 errors of uncorrected composition. The individual-center spectra show the forward construction behind this agreement. The full population is displayed as a spectral map; representative spectra identify the contributions from both parents and all four orientations; and orientation-weighted sums reconstruct the clean ensemble response. These views connect a quantitative prediction to the local objects that generate it. For NV magnetometry and waveguide-integrated sensing, the practical outcome is an explicit model that retains the source operator expressions while calibrating target spectral shifts, disorder, field projection, and optical readout. This use of local spectral information connects with high-resolution NV spectroscopy and ensemble ODMR strain characterization [10,15,39]. Measurements across specimens under controlled field vectors would provide a direct setting for comparing inherited and target-only models, estimating predictive uncertainty, and testing alternative disorder distributions or line-shape kernels. The same composition procedure would then relate those model choices to both the aggregate spectrum and its individual-center contributions.

## VII. CONCLUSION

We connect single-center Hamiltonians to calibrated NV ensemble spectra through explicit individual-center spectral contributions. The pipeline selects effective local Hamiltonians from single-center ODMR measurements, freezes their operator representations, and calibrates their composition to an ensemble response. The zero-field experimental ensemble shows a 56.2% reduction in held-frequency error relative to matched uncorrected composition. Controlled 128-site QuTiP ensembles further predict a wholly withheld magnetic-field condition at errors close to the imposed noise scale. Resolving the simulated spectra into local contributions shows how the calibrated single-center structure builds the collective line shape.

The resulting model retains the inherited spin terms, the target-specific corrections, and the weights that connect local responses to ensemble readout. Within the chosen operator family, these elements provide a defined route for recalibrating a specified sample and field configuration while retaining the source expressions. The independent-site construction treats local spectral variability and additive optical response explicitly, with ensemble dephasing studies providing a broader context for the role of local environments [22]. This connection makes the modeled local contributions a physical basis for interpreting changes in the collective line shape and applying single-center spin knowledge to ensemble characterization.

## DATA AVAILABILITY

The experimental measurements are publicly accessible in the Cardiff University deposit [32]. The study's reproducibility files contain the processed spectra, frozen source models, fitted parameters, numerical replay records, and figure-generation scripts. The retained archive is NV_PRApplied_Source_and_Reproducibility_20261009.zip, with the scientific subarchive Single_to_Ensemble_NV_Pathway_Reproducibility_20261009.zip. The revision adds the individual-center spectral arrays, composition checks, and Blender scene and rendering scripts used for the updated figures. Experimental and simulated data are identified in the accompanying data index.

## ACKNOWLEDGMENTS AND AI-USE DISCLOSURE

OpenAI Codex assisted with manuscript revision, code review, and preparation of the plotting and Blender scripts. The original experimental sources and numerical methods are identified in the text and reproducibility files.

# APPENDIX

## APPENDIX A: EFFECTIVE LOCAL HAMILTONIAN AND NV GEOMETRY

The effective Hilbert space contains a spin-1 electron, spin-1 host nitrogen, and an auxiliary spin-1/2, yielding 18 basis states. The spin matrices follow the QuTiP projection ordering from positive to negative magnetic quantum number. We express frequencies in MHz and magnetic fields in mT. The Hilbert-space dimension is

$$d = 3 \times 3 \times 2 = 18. \tag{A1}$$

With dimensionless spin operators and the Hamiltonian divided by Planck's constant, the specified local family is [13,14]

$$\mathcal{K} = DS_z^2 - 4.96I_z^2 + \sum_{j=1}^{6} z_j c_j O_j + \sum_{k\in\{x,y,z\}} B_k \left(28.02495S_k - 0.003077I_k\right). \tag{A2}$$

The library comprises the following six optional Hermitian terms, each selected by a binary inclusion indicator:

$$(O_1, O_2, O_3) = \left(S_x^2 - S_y^2,\ -S_zI_z,\ -S_xI_x - S_yI_y\right). \tag{A2a}$$

$$(O_4, O_5, O_6) = \left(S_zJ_z,\ S_xJ_x + S_yJ_y,\ S_z\right), \quad z_j \in \{0,1\}. \tag{A2b}$$

The six operators are Hermitian in the complete electron–nitrogen–auxiliary-spin basis. The nitrogen quadrupole term is fixed and the auxiliary nuclear Zeeman term is omitted. The transverse term follows the standard effective NV spin representation [16].

The Hamiltonian is expressed in frequency units, $K = H/h$, with eigenvalues and couplings in MHz. Magnetic-field components are in mT, and dimensionless spin matrices have the projections $m_s$, $m_I$, and $m_J$. The computational tensor order is electron, nitrogen, and auxiliary spin:

$$S_\alpha = s_\alpha \otimes \mathbb{I}_3 \otimes \mathbb{I}_2, \qquad I_\alpha = \mathbb{I}_3 \otimes i_\alpha \otimes \mathbb{I}_2,$$
$$J_\alpha = \mathbb{I}_3 \otimes \mathbb{I}_3 \otimes j_\alpha.$$

Here $s_\alpha, i_\alpha$ are spin-one matrices and $j_\alpha = \sigma_\alpha/2$. In the ordered bases $(+1,0,-1)$ and $\left(+\frac{1}{2}, -\frac{1}{2}\right)$, they are specified by $s_z|m\rangle = m|m\rangle$, $s_+|m\rangle = \sqrt{s(s+1) - m(m+1)}|m+1\rangle$, $s_x = (s_+ + s_-)/2$, and $s_y = (s_+ - s_-)/(2i)$, with the same construction for the other spins. With identities implicit through these definitions, the implemented $18 \times 18$ operator is

$$\begin{aligned}\mathcal{K} = DS_z^2 + QI_z^2 + \gamma_e \mathbf{B}\cdot\mathbf{S} - \gamma_N \mathbf{B}\cdot\mathbf{I} \\ +E\left(S_x^2 - S_y^2\right) - a_\parallel S_zI_z - a_\perp\left(S_xI_x + S_yI_y\right) \\ +k_\parallel S_zJ_z + k_\perp\left(S_xJ_x + S_yJ_y\right) + b_{\mathrm{res}}S_z.\end{aligned} \tag{A3}$$

Equation (A3) specifies the implemented Kronecker-product Hamiltonian. The fixed coefficients are $Q = -4.96$ MHz, $\gamma_e = 28.02495$ MHz/mT, and $\gamma_N = 0.003077$ MHz/mT. The electron Zeeman term uses a positive gyromagnetic-ratio magnitude, whereas the nitrogen Zeeman contribution has a negative sign. For t in microseconds, the frequency-unit propagator is $\exp(-i2\pi Kt)$.

The six coefficients $(E, a_\parallel, a_\perp, k_\parallel, k_\perp, b_{\mathrm{res}})$ define the optional operator library: each is fitted when selected and set to zero otherwise. The field is supplied externally to this module, and $b_{\mathrm{res}}$ is an additional fitted axial frequency offset. Its electronic action overlaps that of $\gamma_e B_z$, so both enter the same axial spin term. Fitted $D$ and $E$ describe effective resonance and transverse-splitting parameters for the local environment [1,9,12].

TABLE I. Six-operator Hermitian library, with effective full-data coefficients for single-center parents A and B and the fitted coefficient bounds. All coefficients are in MHz. A zero entry denotes a term omitted from that parent expression.

| Term | Hermitian operator | Parent A (MHz) | Parent B (MHz) | Bounds (MHz) |
|---|---|---|---|---|
| $E$ | $S_x^2 - S_y^2$ | 1.30245 | 1.06543 | 0–8 |
| $a_\parallel$ | $-S_zI_z$ | 2.23518 | 2.12603 | 0–4 |
| $a_\perp$ | $-\left(S_xI_x + S_yI_y\right)$ | 0.22509 | 5.00000 | 0–5 |
| $k_\parallel$ | $S_zJ_z$ | 0.53017 | 6.64016 | 0–10 |
| $k_\perp$ | $S_xJ_x + S_yJ_y$ | 3.98056 | 0 | 0–5 |
| $b_{\mathrm{res}}$ | $S_z$ | 0.75839 | 0.89182 | 0–4 |

For parent A, the fitted zero-field splitting is 2871.09609 MHz and all six coefficients are retained. For parent B it is 2871.33517 MHz and the fifth coefficient is absent. In operator order, the effective coefficient vectors are

$$\mathbf{c}_A = (1.30245, 2.23518, 0.22509, 0.53017, 3.98056, 0.75839)\ \text{MHz}. \tag{A4}$$

$$\mathbf{c}_B = (1.06543, 2.12603, 5.00000, 6.64016, 0, 0.89182)\ \text{MHz}. \tag{A5}$$

The parent-B transverse nitrogen coefficient reaches its imposed 5 MHz upper bound. The coefficients are retained together with their selected supports and bounds, so the effective source representation is fully specified. Figure 6 summarizes the fitted values and the support variation across frequency folds.

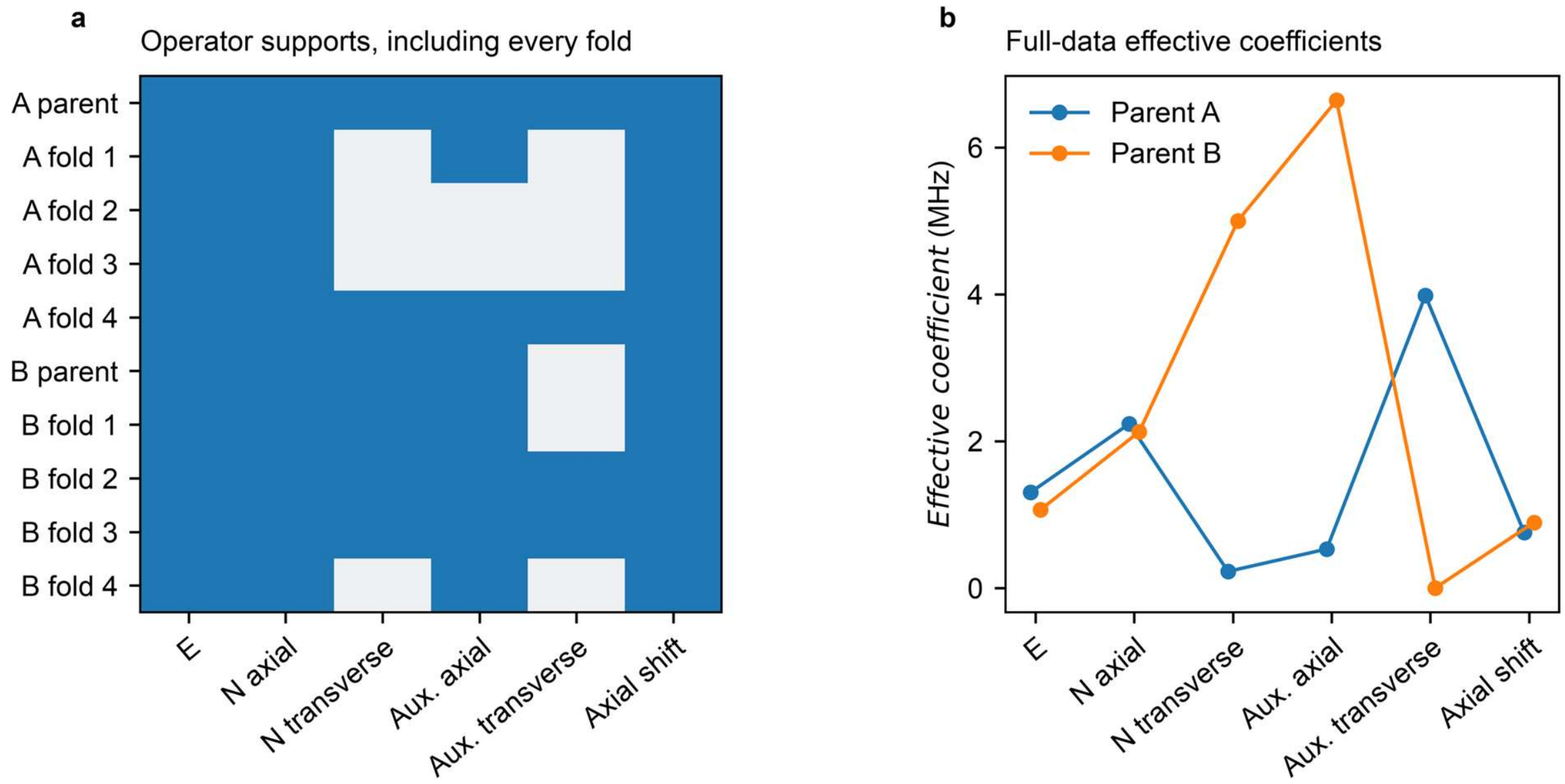

**FIG. 6.** Local expression variability and retained effective coefficients. (a) Operator supports for each full-data source parent and every held-frequency fold. Selected terms are blue and omitted terms light gray. The auxiliary spin basis remains present when its coupling terms are omitted. (b) Full-data coefficients in the stated operator order, in MHz. The parent-B transverse nitrogen coefficient reaches its 5 MHz upper bound. These effective coefficients and their selected supports define the frozen source models inherited by the target workflow.

### A. Auxiliary-spin coupling and physical motivation

The auxiliary spin-1/2 represents a nearby nuclear spin in the effective local model. The source experiment reports additional hyperfine structure of order 6 MHz [31], consistent with nearby-$^{13}$C spectroscopy [40]. The selected coupling terms describe this additional resolved structure.

$$\mathcal{K}_C = \sum_{\alpha,\beta} S_\alpha A^{(C)}_{\alpha\beta} J_\beta - \gamma_C \mathbf{B}\cdot\mathbf{J}. \tag{A6}$$

For a nearby $^{13}$C nucleus, $\gamma_C$ denotes the nuclear gyromagnetic ratio. The effective hyperfine model restricts the coupling tensor to $diag(k_\perp, k_\perp, k_\parallel)$ in the NV frame and omits the $^{13}$C Zeeman term. The auxiliary spin has no quadrupole contribution for $J = 1/2$. With $k_\parallel = k_\perp = 0$, the auxiliary spin doubles the degeneracy of the nine-level electron–nitrogen model.

### B. Local frames and NV orientations

The local z axis follows the NV symmetry axis. A transverse splitting term refers to axes in the plane perpendicular to that local axis. Transforming laboratory magnetic fields and strain tensors into the local frame requires the full orientation matrix, including its transverse-axis convention.

We choose laboratory x, y, and z parallel to the diamond [100], [010], and [001] axes, respectively.

The four tetrahedral NV directions are

$$n_q \in \frac{1}{\sqrt{3}}\{(1,1,1),\ (1,-1,-1),\ (-1,1,-1),\ (-1,-1,1)\}. \tag{A7}$$

Define $\mathbf{e}_{z,q} = \mathbf{n}_q$, $\mathbf{e}_{x,q} = (\mathbf{e}_Z \times \mathbf{n}_q)/|\mathbf{e}_Z \times \mathbf{n}_q|$ and $\mathbf{e}_{y,q} = \mathbf{n}_q \times \mathbf{e}_{x,q}$.

With $R_q$ having these local unit vectors as rows, $\mathbf{B}_q = R_q \mathbf{B}_{\text{lab}}$.

A second-rank strain tensor is transformed into the local frame as $\boldsymbol{\epsilon}_q = R_q \boldsymbol{\epsilon}_{\text{lab}} R_q^{\mathsf{T}}$ ; similar coordinate transformations are used in waveguide NV calculations [39].

For an arbitrary laboratory-to-crystal alignment, the field and strain transformations include the corresponding mounting rotation. Axial and transverse components are defined relative to the individual NV axis.

The single-center fits use a magnetic field along the local z axis. The ensemble calculation includes all four tetrahedral orientations explicitly, with Gaussian quadrature describing the distribution of local splitting offsets.

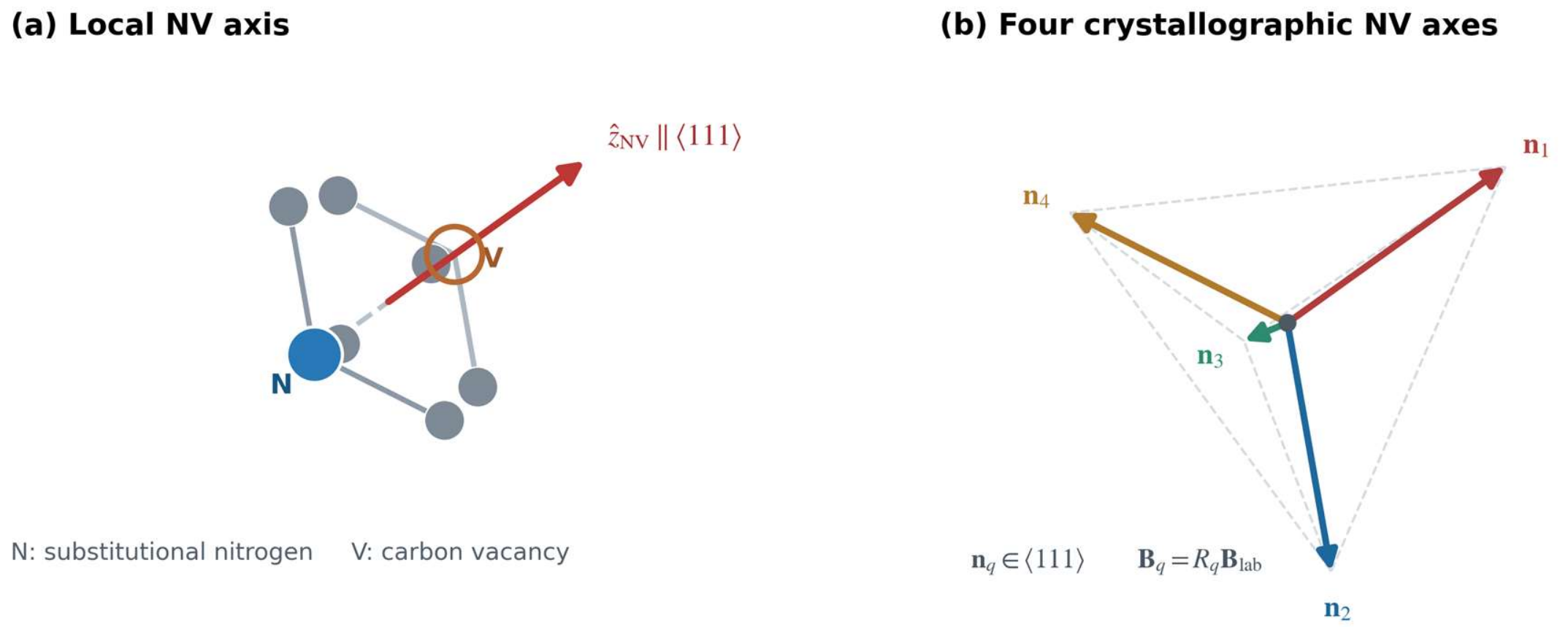

**FIG. 7.** Author-reconstructed schematic of local coordinate system and crystallographic orientations. (a) Schematic nitrogen–vacancy pair defining the local NV symmetry axis. (b) Four tetrahedral NV orientations in the diamond crystal frame, with vectors expressed in the crystal-coordinate system. The schematic provides the local axis convention used to transform lab-frame fields in the ensemble model.

## APPENDIX B: SPECTRAL SURROGATE AND SYMBOLIC SELECTION

Diagonalization of the local Hamiltonian supplies the eigenfrequencies and eigenvectors. We define brightness using the projector onto the electronic $m_s = 0$ subspace and partition states at an expectation value of one-half. Allowed transitions are assigned phenomenological oscillator weights from transverse electronic spin matrix elements and brightness contrast; only positive weights greater than $10^{-9}$ are retained. Each transition contributes a line kernel

$$L(f;\nu,\Gamma) = \frac{1}{1+4(f-\nu)^2/\Gamma^2}. \tag{B1}$$

Transition weights are normalized separately in the two frequency groups, defined about the midpoint of the full transition range. The resulting summed Lorentzian response serves as the phenomenological line-shape model for both continuous-wave and pulsed spectra.

For each single-center acquisition, the two response components and the optical offset and background slope enter linearly. Their coefficients are fitted within the nonlinear parameter search by variable projection [36,37]. For an ensemble, the corresponding template, offset, and normalized-frequency slope give

$$\hat{\boldsymbol{y}} = X(\boldsymbol{\theta})\boldsymbol{\beta}, \hat{\boldsymbol{\beta}} = \arg min_{\boldsymbol{\beta}\in\mathcal{B}} \|X(\boldsymbol{\theta})\boldsymbol{\beta} - \boldsymbol{y}\|_2^2. \tag{B2}$$

Linear readout coefficients are bounded. For a single-center acquisition, the two nonnegative contrast coefficients lie between zero and 0.7, the optical offset between 0.7 and 1.3, and the frequency-background slope between −0.02 and 0.02 $MHz^{-1}$ relative to 2870 MHz. Ensemble-specific bounds and normalization follow the archived numerical implementation.

The discrete expression and continuous parameters are scored using a Bayesian-information-criterion form [35]

$$C = n\ln\left[\max\left(\frac{\mathrm{SSE}}{n}, 10^{-20}\right)\right] + k\ln n \tag{B3}$$

The score includes nonlinear parameters, linear readout coefficients, and one estimated residual-variance parameter. The six-bit grammar defines 64 supports; genetic search evaluates 24 distinct expressions over three generations and a population of six. Coefficients use two bounded initializations in each fold. Source parents are the stored seed-11 full-data outputs. The all-six-term benchmark uses the same coefficient-initialization budget.

The zero-field splitting is bounded to 2866–2877 MHz, with optional coefficient bounds of 0–8, 0–4, 0–5, 0–10, 0–5, and 0–4 MHz in operator order. Linewidth lies between 0.15 and 12 MHz.

## APPENDIX C: RAW DATA AND WITHHELD-FREQUENCY SPLITS

The experimental data accompany Guo et al. [31,32]. Reprocessing the archived raw workbooks reproduced 24 processed spin-data arrays; the source archive is 847809 bytes with a matching recorded MD5 checksum. This analysis uses four single-center ODMR traces and two ensemble traces. Figure 8 displays the observations and frequency-fold assignments.

For each frequency-sorted single-center or zero-field ensemble acquisition, consecutive four-point blocks are allocated cyclically among four folds. For the finite-field ensemble, the block size is five. The fold rule is

$$g(i;b) = \left\lfloor \frac{i}{b} \right\rfloor mod\ 4, b = 4 \text{ or } 5. \tag{C1}$$

Each finite-field fold contains 65 observations. Every point is predicted from a fit excluding its assigned fold, and the common Hamiltonian for each center is fitted across its CW and pulsed acquisitions. The pooled error over withheld observations is

$$\mathrm{RMSE}_{\mathrm{pooled}} = \sqrt{\frac{\sum_i (\hat{y}_i - y_i)^2}{N}}. \tag{C2}$$

The experimental data contain 87 single-A, 210 single-B, 38 zero-field ensemble, and 260 finite-field ensemble observations. The reported cross-validation measures interpolation at held frequencies within each scan.

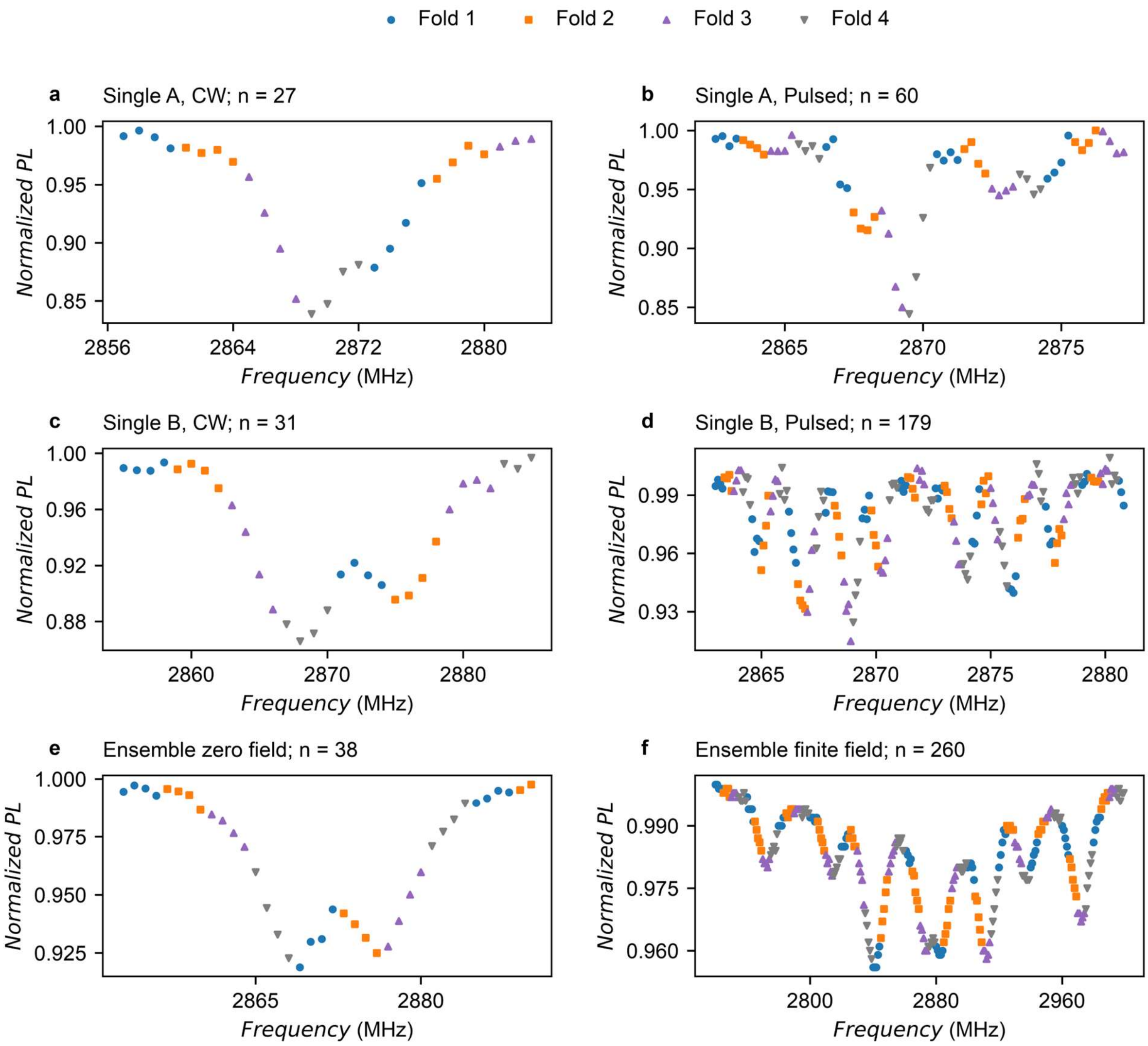

**FIG. 8.** Experimental observations and validation partitions. (a–d) CW and pulsed traces for the two source centers. (e,f) Zero-field and finite-field ensemble traces. Color and marker shape identify four cyclic folds. Single-center and zero-field target scans use consecutive groups of four sorted-frequency observations; the finite-field target uses groups of five and has 65 observations per fold. Each observation is predicted once with its own fold withheld. The original sampling densities are retained, and normalized PL is the processed measurement quantity. The four colors identify frequency partitions within each acquisition.

## APPENDIX D: ENSEMBLE CALIBRATION AND FIELD TRANSFER

The target correction modifies the inherited mean zero-field splitting, transverse coefficient, and effective axial shift; optional Gaussian disorder broadens the local splitting distribution. The source hyperfine couplings remain frozen. Symbolically,

$$\frac{H_t}{h} = \frac{H_p}{h} + (\delta D + u)S_z^2 + \delta E(S_x^2 - S_y^2) + \delta b_z S_z. \tag{D1}$$

Five-node Gaussian quadrature approximates the splitting-offset distribution, whose standard deviation is constrained to be nonnegative. Parent weights are nonnegative and sum to unity. Six binary choices encode inclusion of parent A, parent B, splitting shift, transverse shift, splitting disorder, and axial correction. Candidates containing no source parent are invalid; their failure sentinels remain in the archived search log.

The primary full-data expression (seed 11) selects parent A, with corrections of 1.34785 MHz to D, 2.01934 MHz to E, and 0.92153 MHz to the effective axial term. Linewidth is 6.77652 MHz, and Gaussian splitting disorder is zero. Seeds 23 and 37 select a parent mixture with weight 0.54904 for A and full-data RMSE 0.0018444, compared with 0.0019254 for seed 11.

At finite field, source parents and the primary zero-field Hamiltonian corrections remain fixed. Field direction, linewidth, four orientation contrasts, baseline, and spectral slope are fitted. The reported 5.17 mT magnitude gives a held-frequency RMSE of 0.0081991, whereas a freely fitted effective magnitude gives 0.0032094. The four fitted effective magnitudes span 3.7416–3.7741 mT.

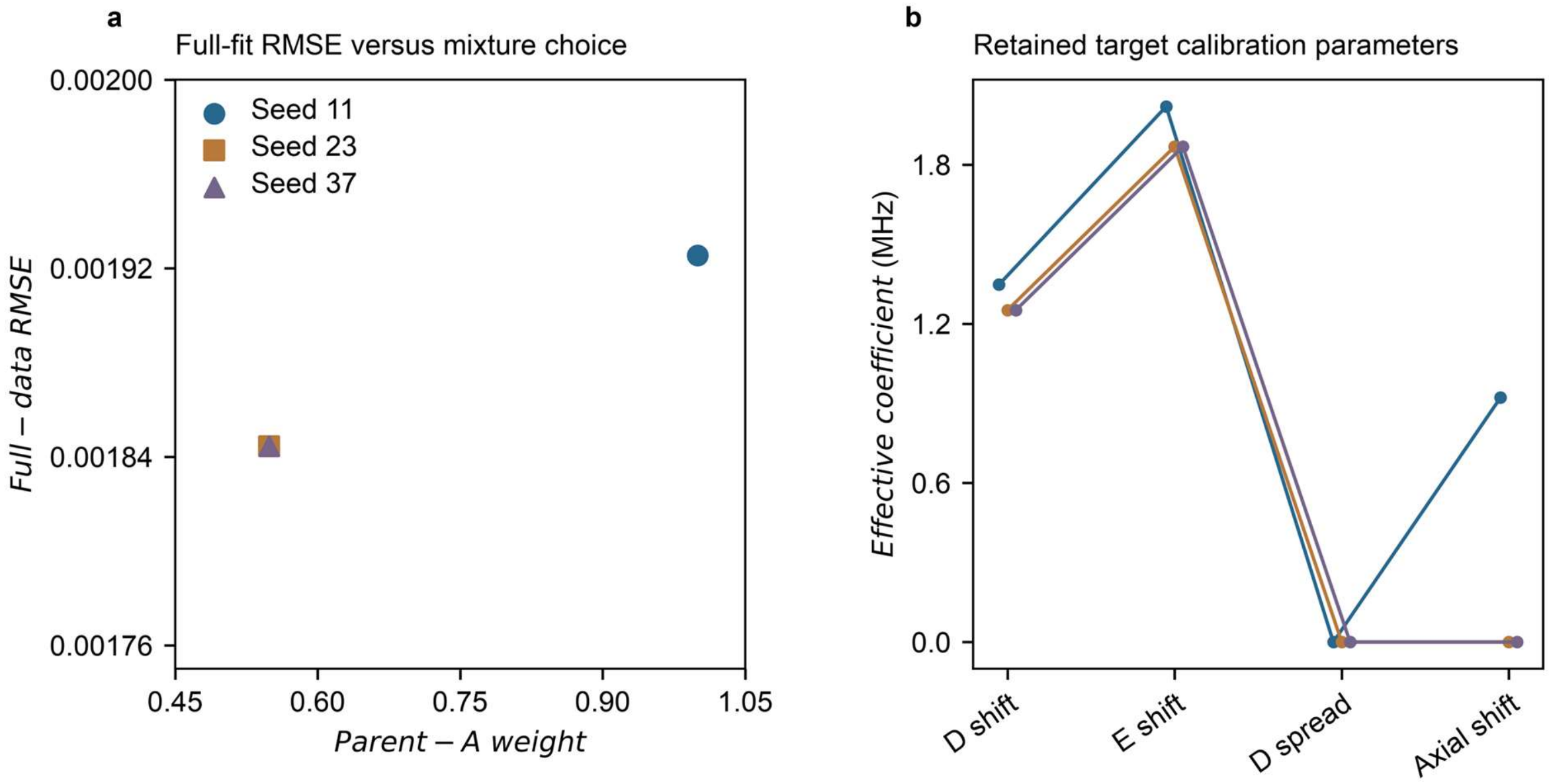

**FIG. 9.** Target expression and coefficient variation across search seeds. (a) Full-data zero-field target RMSE versus parent-A weight for seeds 11, 23, and 37. Seeds 23 and 37 coincide and remain distinct records; seed 11 is the prespecified primary expression. (b) Splitting shift, transverse shift, splitting-distribution standard deviation, and effective axial shift for each seed, in MHz. Similar aggregate errors accompany different parent choices and correction supports. These full-data solutions describe the variation among the retained effective ensemble models.

## APPENDIX E: INDEPENDENT-SITE AVERAGING AND QUTIP CONSTRUCTION

For noninteracting NV sites, the formal ensemble Hamiltonian is a sum of local operators acting on separate tensor factors:

$$H_{\text{ens}} = \sum_{r=1}^{M} I^{\otimes(r-1)} \otimes H_r \otimes I^{\otimes(M-r)}. \tag{E1}$$

Because operators on different sites commute, factorization of the time evolution follows for an initial product state:

$$U(t) = \otimes_{r=1}^{M} \exp(-i2\pi H_r t/h), \rho(0) = \otimes_{r=1}^{M} \rho_r(0). \tag{E2}$$

The factor $2\pi$ converts the frequency-unit Hamiltonian to angular frequency when time is measured in microseconds. This factorization explains why local observables can be assembled without propagating the full tensor-product state. For an additive observable with nonnegative normalized site weights, the expectation value reduces to a weighted sum of local expectations:

$$\langle A\rangle_t = \sum_{r=1}^{M} w_r\, Tr\left[A_r U_r(t)\rho_r(0)U_r^{\dagger}(t)\right], \sum_r w_r = 1. \tag{E3}$$

The numerical implementation evaluates stationary eigenstate spectra and combines their line responses according to

$$S_{\mathrm{ens}}(f) = \sum_o w_o \sum_p w_p \int S_p(f; R_o\boldsymbol{B}, u)\rho(u)du. \tag{E4}$$

For independent sites with additive optical readout, the normalized ensemble response is a weighted average of local spectra. Nonnegative orientation and parent weights sum to unity, and a finite population replaces the continuous offset distribution by its empirical sample. Figures 5(g)–6(i) evaluate that finite sum using the saved site identities, orientations, and offsets. Figure 10 presents the local and explicit two-site Hamiltonian matrices used to check the additive tensor construction.

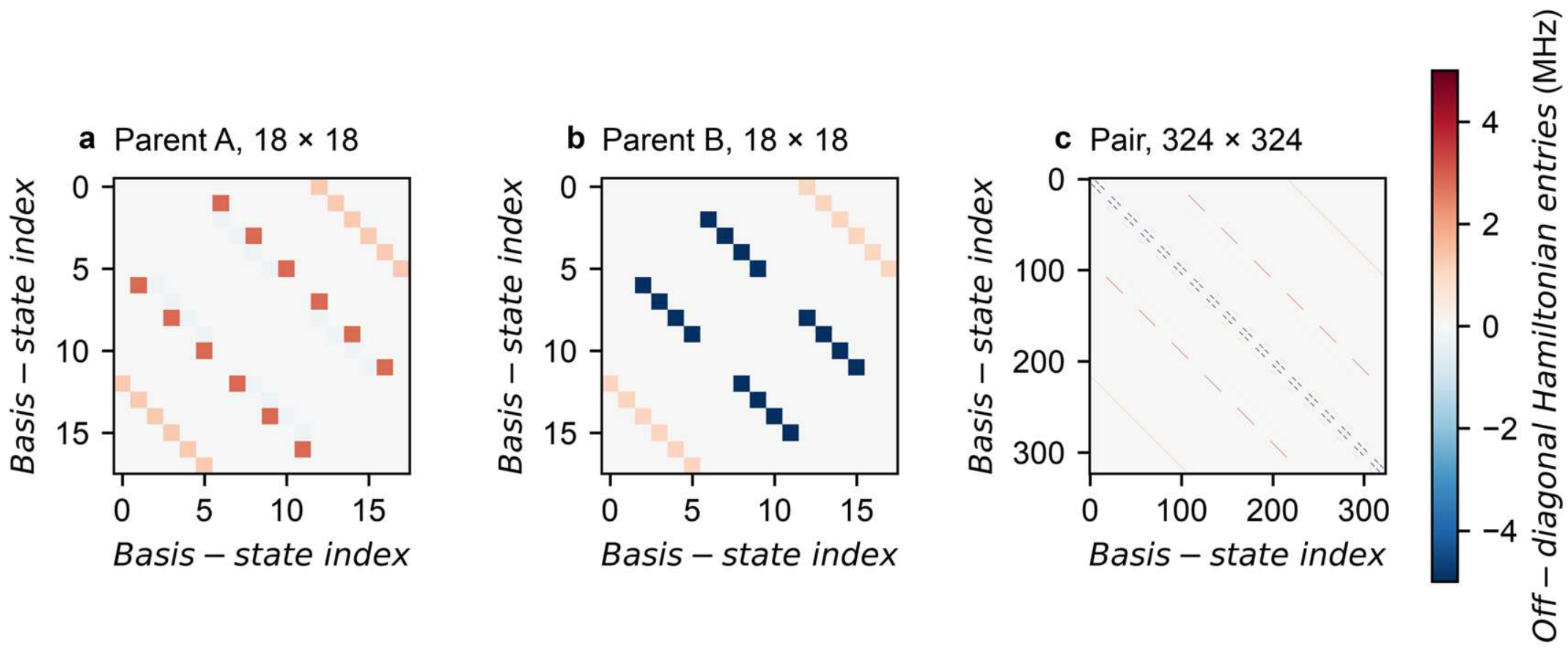

**FIG. 10.** Numerical local and two-site Hamiltonian structure. (a,b) Retained 18-dimensional parent-A and parent-B matrices. (c) Explicit additive two-site Hamiltonian of dimension 324. Diagonal entries are removed only in this display to expose smaller couplings; the common signed color scale reports real off-diagonal entries in MHz. The companion NPZ data retain the full matrices, including their diagonals. Basis indices follow the saved tensor ordering. These matrices and the corresponding zero-difference checks verify the numerical implementation of the independent-site tensor sum.

Each seed contains 128 sites: 32 per tetrahedral NV orientation and 16 from each source parent in each orientation. Local splitting offsets are sampled with standard deviation 0.7 MHz and then centered on their empirical mean. The generator specifies mean D and E corrections of 1.3 and 1.6 MHz, linewidth 3.5 MHz, equal parent weights, contrast 0.18, and optical baseline 1. The three finite-site offset standard deviations are approximately 0.654, 0.755, and 0.663 MHz.

The laboratory field direction is proportional to the vector with components 0.31, 0.47, and 0.826 and is normalized before use. Spectra contain 321 frequencies from 2710 to 3030 MHz. The generator uses QuTiP tensor spin operators [33,34], while the learner constructs the operators independently with NumPy. Both use the specified operator library and optical transition rule. QuTiP evaluates 1152 local eigensystems across the three seeds and three fields. An explicit 324-dimensional two-site matrix separately checks the additive tensor construction from two local Hamiltonians.

The learner fits five continuous parameters: linewidth, two mean corrections, splitting spread, and source-parent fraction. Four binary indicators select the optional corrections; the search evaluates up to 12 expressions using two continuous starts. Generator baseline and optical contrast are held fixed and known to the learner. Selection and fitting use only the 0 and 3 mT conditions. Figure 11 reports residuals for all conditions and seeds.

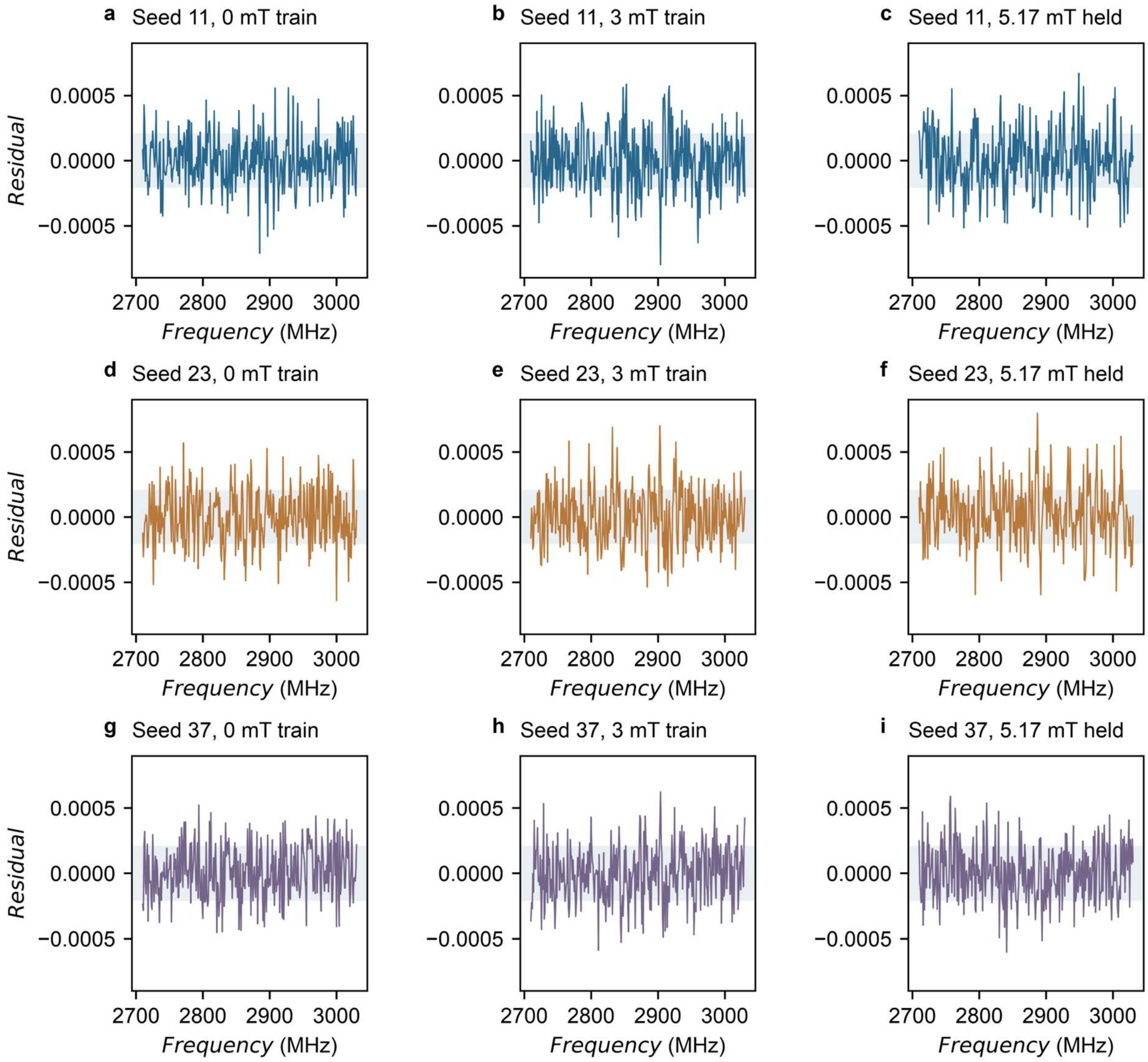

**FIG. 11.** Simulation residuals for every field and seed. Rows correspond to seeds 11, 23, and 37; columns to 0, 3, and 5.17 mT. The first two fields train expression selection and coefficient fitting, and the final condition is withheld entirely. Each panel contains all 321 frequencies with common residual limits of −0.0009 to 0.0009. Residuals are prediction minus noisy finite-ensemble observation in normalized PL. Shading marks ±0.000200, the standard deviation of the added noise. Displaying the training and held conditions together shows the change in residual structure across the field prediction.

## APPENDIX F: SEARCH TRAJECTORIES AND REPRODUCIBILITY

Figure 12 uses a display-only min–max normalization of valid running-best training scores, with bounds common to the three seeds for each task:

$$\tilde{C} = \frac{C - C_{min}}{C_{max} - C_{min}}. \tag{F1}$$

For each task, scores are normalized using the full range of valid running-best records across three seeds. Smaller values indicate better scores, with the minimum mapped to zero. The plots display 105 valid running-best points; three invalid candidates remain documented in the archived logs.

The simulation search is initialized with the empty expression, the full 4-correction expression, and 4 single-correction expressions. The full expression is selected at evaluation two for all seeds. Saved nonlinear evaluation counts for the selected initializations are 8, 6, and 8, respectively.

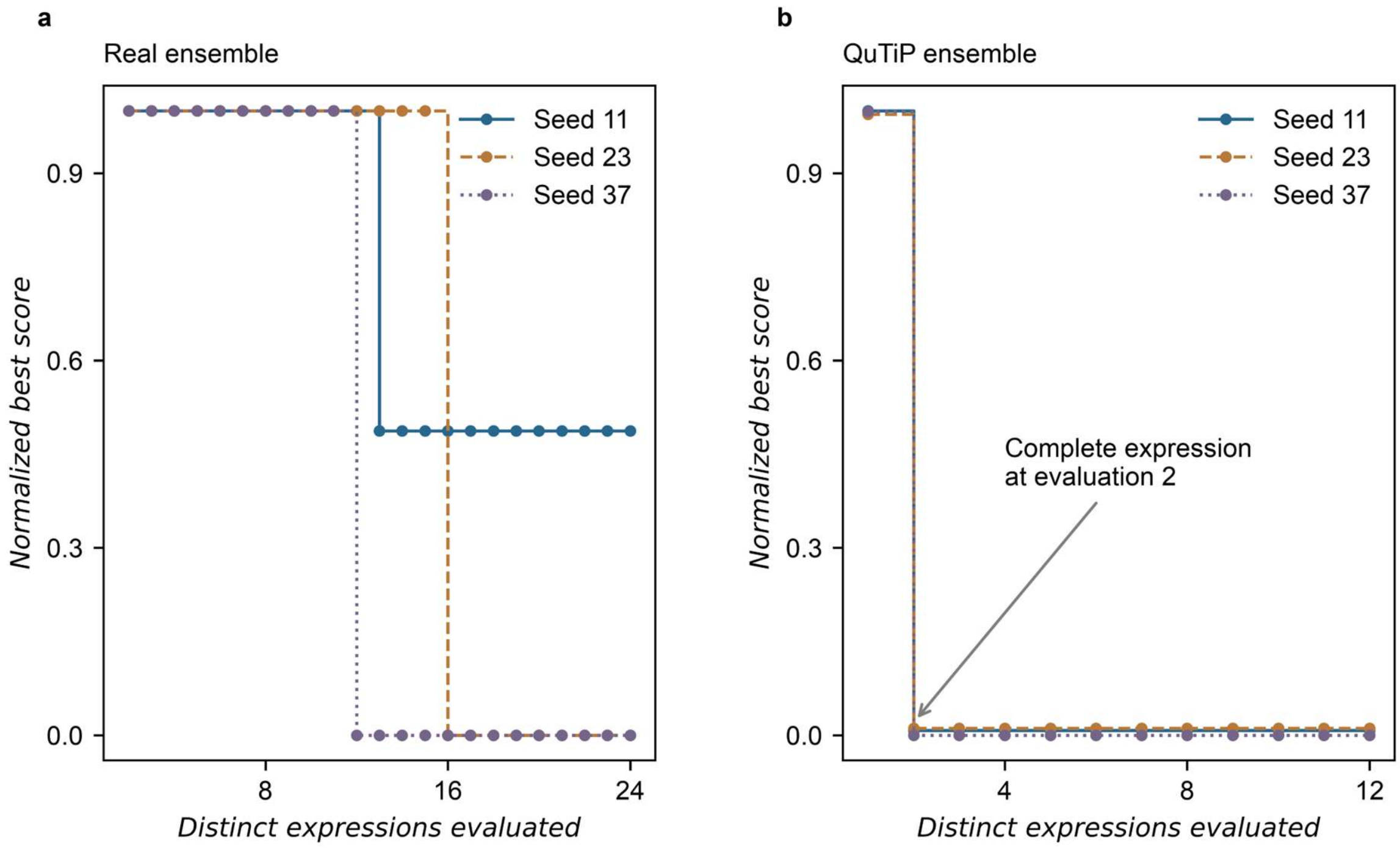

**FIG. 12.** Display-normalized expression-search trajectories. (a) Experimental zero-field target search. (b) Simulated ensemble search. Valid running-best scores from all three seeds are mapped to the interval zero to one using a common minimum and maximum within each task. Lower values indicate better scores, and zero denotes the best observed valid training score for that task. All 105 valid running-best records are retained. Three invalid experimental candidates are recorded as failures in the original logs and excluded from this normalization. The full simulation expression is an initialized candidate and is selected at evaluation two; its continuous coefficients are fitted to the training spectra.

An isolated replay reproduced the 24 archived numerical outputs within the reported tolerances, including experimental fold fits, the ensemble calibration, simulated field prediction, and matrix-consistency calculations. The historical source parents and search records are preserved with the reproducibility archive.